# Charge, geometry, and effective mass

Gerald E. Marsh

Argonne National Laboratory (Ret)
5433 East View Park
Chicago, IL 60615

E-mail: geraldemarsh63@yahoo.com

**Abstract.** Charge, like mass in Newtonian mechanics, is an irreducible element of electromagnetic theory that must be introduced *ab initio*. Its origin is not properly a part of the theory. Fields are then defined in terms of forces on either masses—in the case of Newtonian mechanics, or charges in the case of electromagnetism. General Relativity changed our way of thinking about the gravitational field by replacing the concept of a force field with the curvature of space-time. Mass, however, remained an irreducible element. It is shown here that the Reissner-Nordström solution to the Einstein field equations tells us that charge, like mass, has a unique space-time signature.

PACS: 04.20.Cv, 04.20.Dw.

Key Words: Reissner-Nordström, charge, curvature.

*Charge, geometry, and effective mass*

**Introduction.**

The Reissner-Nordström solution is the unique, asymptotically flat, and static solution to the spherically symmetric Einstein-Maxwell field equations. Its accepted interpretation is that of a charged mass characterized by two parameters, the mass $M$ and the charge $q$. While this solution [1] has been known since 1916, there still remains a good deal to be learned from it about the nature of charge and its effect on space-time.

It will be shown here that if the source of the field is the singularity of the vacuum Reissner-Nordström solution, only the Schwarzshild mass is seen at infinity, with the charge and its electric field making no contribution. In particular, if the charge alone is the source of the field, the effective mass seen at infinity vanishes. This is not the case when the source of the field is a "realistic" source characterized by a mass and proper charge density [5]. It will also be seen that the presence of charge results in a negative curvature of space-time.

The Reissner-Nordström solution is given by

$$ds^2 = -\left(1 - \frac{2m}{r} + \frac{Q^2}{r^2}\right)dt^2 + \left(1 - \frac{2m}{r} + \frac{Q^2}{r^2}\right)^{-1} dr^2 + r^2(d\theta^2 + \sin^2\theta\, d\phi^2), \quad (1)$$

where $m = GM/c^2$ and $Q = (G^{1/2}/c^2)\, q$. The Reissner-Nordström metric reduces to that of Schwarzschild for the case where $Q = 0$. Notice that this metric takes the Minkowski form when $r = Q^2/2m$.

If $Q^2 < m^2$, this solution has two apparently singular surfaces located at $r_\pm = m \pm \sqrt{m^2 - Q^2}$. These are coordinate singularities that may be removed by choosing suitable coordinates and extending the manifold. If $Q^2 = m^2$, these surfaces coalesce into a single surface located at $r = m$, and if $Q^2 > m^2$ the metric is non-singular everywhere except for the origin. These singular surfaces play no role in what follows. An extensive discussion of the vacuum Reissner-Nordström and Schwarzschild solutions, along with their Penrose diagrams is given in Hawking and Ellis [2].

Most applications of the Reissner-Nordström solution would be outside a body responsible for the charge and mass. Here it is the *vacuum* solution to the field equations, considered to be valid for all values of *r*, that is of interest.

Like the vacuum Schwarzschild solution, the Reissner-Nordström vacuum solution has an irremovable singularity (in the sense that it is not coordinate dependent) at the origin representing the source of the field. In what follows, only the Reissner-Nordström solution having this singularity as a source of the field will be considered.

The interesting thing about the singularity is that it is time-like so that clocks near the singularity run *faster* than those at infinity. It is also known that the singularity of the Reissner-Nordström solution is repulsive in that time-like geodesics will not reach the singularity.

**Curvature in the Reissner-Nordström solution.**

If one computes the Gaussian curvature associated with the Schwarzschild solution it is readily seen that the curvature vanishes. Higher order scalars, such as the Kretschmann scalar given by $K = R \quad R$, do not vanish, but their interpretation is problematic [3]. Of course the curvature of space-time around a Schwarzschild black hole does not vanish since the curvature tensor does not vanish. More important for the present discussion is that a simple way to determine the sign of the curvature is well known.

*Charge, geometry, and effective mass*

Consider first the Schwarzschild solution. Draw a circle on the equatorial plane where $\theta = \pi/2$ centered on the origin. The circumference of this circle is $2\pi r$. The proper radius from the origin to the circle is given by

$$\int_0^r \sqrt{g_{11}}\, dr = \int_0^r \left(1 - \frac{2m}{r}\right)^{-\frac{1}{2}} dr \geq r.$$
(2)

Consequently, the ratio of the circumference of the circle to the proper radius is less than or equal to $2\pi$. This tells us that the space is positively curved. Now consider a *negative* mass. The inequality sign in Eq. (2) reverses so that the ratio of the circumference of a circle to its proper radius is greater than $2\pi$—with the conclusion that the space surrounding a negative mass has a negative curvature.

The case of the Reissner-Nordström solution is more interesting. Setting

$$g_{00} = -\left(1 - \frac{2m}{r} + \frac{Q^2}{r^2}\right) \quad \text{and} \quad g_{11} = \left(1 - \frac{2m}{r} + \frac{Q^2}{r^2}\right)^{-1},$$

and using the above method of determining the spatial curvature gives the results shown in Table 1. For $r < Q^2/2m$, one has a *negatively* curved space-time, which is embedded in a positively curved space-time with a 2+1 dimensional boundary having the Minkowski form between them. In the region between the time-like singularity at the origin and the 2+1 dimensional hypersurface, the space-time is negatively curved independent of the sign of the charge. This implies that charge manifests itself as a negative curvature—just as mass causes a positive curvature.

|  | $r > Q^2/2m$ | $r = Q^2/2m$ | $r < Q^2/2m$ |
|---|---|---|---|
| $g_{00}$ | $-1$ | $-1$ | $<-1$ |
| $g_{11}$ | $>1$ | $1$ | $<1$ |
| Spatial Curvature | Positive | Flat | Negative |

Table 1. The metric coefficients $g_{00}$ and $g_{11}$ for different ranges of $r$, and the sign of the spatial curvature in these regions.

That charge effectively acts as a negative mass can also be seen from the equations governing the motion of a test particle near a Reissner-Nordström singularity. For an uncharged particle falling inward towards the singularity the radial acceleration is [4], [5], [6], [7]

$$\frac{d^2 r}{d\tau^2} = -\frac{1}{r^2}\left(m - \frac{Q^2}{r}\right). \tag{3}$$

The gravitational field that affects the test particle varies with distance from the singularity and becomes repulsive when the effective mass $m_{eff} = \left(m - \frac{Q^2}{r}\right)$ becomes negative at $r < Q^2/m$. Neutral matter falling into the singularity would therefore ultimately accumulate on the 2+1 dimensional spherical hypersurface where $m_{eff} = 0$.

Thus, by means of very straight-forward considerations, the Reissner-Nordström solution leads to the conclusion that charge—of either sign—causes a negative curvature of space-time.

**The electric field**

This section is devoted to a general relativistic calculation of the effective mass of the vacuum Reissner-Nordström solution: first, of that contained within the interior of a spherical surface of radius $R$, centered on the singularity—and designated $M_{Eff}^{In}$; and second, the effective mass of the electric field alone outside that surface—designated $M_{Eff}^{Out}$. The key references for what follows are Synge [8] and Gautreau and Hoffman [9].

Synge gives the following Stokes relation* for a 3-dimensional volume, $v_3$, bounded by a closed 2-surface $v_2$:

$$\oint_{v_2} V_{,i} n^i dv_2 = \frac{1}{2} \int_{v_3} \left( G_4^4 - G_i^i \right) V \, dv_3. \tag{4}$$

Here, $dv_2$ and $dv_3$ are the invariant elements of area and volume, $G$ is the Einstein tensor, and $V$ is defined by the line element

$$ds^2 = g_{ij} dx^i dx^j - V^2 dt^2, \tag{5}$$

which, at infinity, is assumed to take the form of the Minkowski metric. $n^i$ is the outward unit normal to the surface $v_2$. Einstein's equations, $G_\mu = -\kappa T_\mu$, with $\kappa = 8\pi$, allow Eq. (4) to be written as

---

* Greek indices take the values 1, 2, 3, 4 and Latin indices 1, 2, 3. To avoid unnecessary confusion, the notation used here is generally consistent with that found in the relevant literature.

*Charge, geometry, and effective mass*

$$\oint_{v_2} V_{,i} n^i dv_2 = 4 \int_{v_3} \left(T_i^i - T_4^4\right) V \, dv_3. \tag{6}$$

The integral on the right hand side of this equation corresponds to the *total effective mass* enclosed by the surface $v_2$. This is known as Whittaker's theorem [10]. Thus,

$$M_{Eff}^{In} = \frac{1}{4} \oint_{v_2} V_{,i} n^i dv_2. \tag{7}$$

Note that the effective mass, as defined by Eqs. (6) and (7), depends only on the energy-momentum tensor and the $g_{00}$ component of the metric. Choose a spherical surface of radius $R$ with the Reissner-Nordström singularity at the origin. From Eq. (1), $V$ is given on the surface by

$$V = \left(1 - \frac{2m}{R} + \frac{Q^2}{R^2}\right)^{\frac{1}{2}}, \tag{8}$$

$dv_2 = R^2 \sin\theta \, d\theta \, d\phi$, and $n^i = (V, 0, 0)$. Substituting into Eq. (7) gives the result quoted above [just after Eq. (3)] for $m_{eff}$ at a distance $R$ from the singularity

$$M_{Eff}^{In} = m - \frac{Q^2}{R}. \tag{9}$$

For asymptotically flat space-times, global quantities such as the total energy can be defined as surface integrals in the asymptotic region. This is the basis for definition of the ADM energy (or mass) [11]. What will be shown here is that for $R \to \infty$, the sum of the effective mass within the surface $v_2$ and that exterior to $v_2$ is the Schwarzschild mass. This is true for the vacuum solution being considered here, not necessarily for realistic sources such as those considered by Cohen and Gautreau [5].

Whittaker's theorem allows the effective mass enclosed by the surface $v_2$, which is composed of the mass located at the origin and that corresponding to the electric field within $v_2$, to be written as in Eq. (9).

One can also compute the effective mass exterior to the surface $v_2$. There, the only the energy density to be found is that associated with the electric field. By summing the effective mass found in the volumes both interior and exterior to $v_2$, one obtains the effective mass enclosed by the surface at infinity; that is, the ADM mass. Given that global quantities defined by surface integrals in the asymptotic region cannot generally be written as volume integrals over the interior region, this is a somewhat surprising result.

How to use the relation of Eq. (6) to compute the electric field energy in the volume *exterior* to the spherical surface of radius $R$ centered on the singularity can be understood by referring to Figure 1. The volume of interest is $v'_3$ *exterior* to the surface $v_2$. It has two boundary components, the "surface at infinity" and $v_2$ itself.

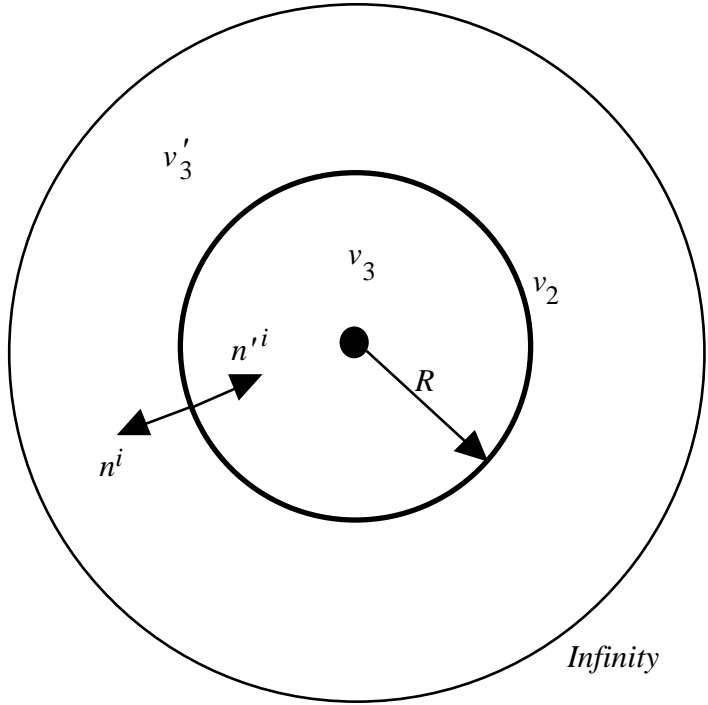

Figure 1. The Reissner-Nordström singularity is located at the center of the spherical surface $v_2$ of radius $R$ enclosing the volume $v_3$. The outward pointing unit normal to $v_2$ is $n^i$. The surface enclosing the volume $v'_3$ is composed of the point at infinity and $v_2$. The outwardly pointing unit normal to $v_2$, when acting as a boundary component of $v'_3$, is $n'^i$.

Since the surface integral at infinity vanishes, Eq. (6) for the volume $v'_3$ may be written as

$$4 \int_{v_3} \left(T_i^i - T_4^4\right) V \, dv_3 = \oint_{v_2} V_{,i} n^i dv_2. \tag{10}$$

The $V$ in Eq. (6) has been changed to $V'$ in Eq. (10). The reason for this is that the energy-momentum tensor in the volume $v'_3$ must be restricted to the contribution from only the electric field since no masses exist in $v'_3$. The way to do this is to recognize that the Reissner-Nordström solution remains a solution to the Einstein field equations even when the mass $m$ is set equal to zero. The resulting metric is that for a massless point charge, which—as discussed above—has a negative curvature and is repulsive. The energy-momentum tensor for the electric field nonetheless has a positive energy density. The $V'$ that should be used in Eq. (10) is therefore that from the metric for a massless point charge; i.e.,

$$V = \left(1 + \frac{Q^2}{r^2}\right)^{\frac{1}{2}}.$$

(11)

Note that $r$ takes the fixed value $R$ when computing the surface integral.

Because $n^i = -n'^i$, the effective mass contained in the volume $v'_3$ exterior to $v_2$ is

$$M_{Eff}^{Out} = \frac{1}{4}\oint_{v_2} V_{,i}\, n'^i dv_2 = -\frac{1}{4}\oint_{v_2} V_{,i}\, n^i dv_2 = -\int_R \left(T_i^i - T_4^4\right) V\, dv_3.$$

(12)

$M_{Eff}^{Out}$ can be evaluated by simply using the second integral in Eq. (12), which was already evaluated for $V$ above.  Taking account of the orientation of the surface and the substitution of $V'$, the result is

$$M_{Eff}^{Out} = \frac{Q^2}{R}.$$

(13)

Combined with Eq. (9), this results in

$$M_{Eff}^{In} + M_{Eff}^{Out} = m.$$

(14)

What this tells us is that the "negative mass" associated with the charge $Q$ [see Eq. (9)] is exactly compensated by the effective mass contained in the electric field present in the volume exterior to the surface $r = R$. If the radius $r = R$    , the effective mass contained within the surface at infinity is $m$, the Schwarzschild or equivalently, the ADM mass.

One can also obtain the result given in Eq. (14) by directly evaluating the last integral on the right hand side of Eq. (12).  This will be done here for the sake of completeness as well as a confirmation of Eq. (13) above.  To begin with, an identity

*Charge, geometry, and effective mass*

relating the energy-momentum tensor $F_{\mu\nu}$ of the electric field to the scalar potential is needed.

If the energy-momentum tensor

$$T_{\mu\nu} = F_{\mu\alpha}F_{\nu}{}^{\alpha} - \frac{1}{4}g_{\mu\nu}F_{\alpha\beta}F^{\alpha\beta}, \tag{15}$$

where

$$F_{\mu\nu} = \frac{\partial A_{\nu}}{\partial x^{\mu}} - \frac{\partial A_{\mu}}{\partial x^{\nu}},$$

is restricted to the case where only electric fields are present, so that

$$A_{\mu} = \left(0, 0, 0, \frac{1}{\sqrt{4\pi}}\right), \tag{16}$$

then it is readily shown that

$$F_{i4} = \frac{1}{\sqrt{4\pi}}\varphi_{,i} \quad \text{and} \quad F^{i4} = -\frac{1}{\sqrt{4\pi}}V^{-2}g^{ij}\varphi_{,j}. \tag{17}$$

Equations (17) allow the energy-momentum tensor to be written as

$$4\pi T_{ij} = \frac{2}{V^2}\left(\frac{1}{2}g_{ij}\Delta_1\varphi - \varphi_{,i}\varphi_{,j}\right), \tag{18}$$

where $\Delta_1\varphi$ is a differential parameter of the first order defined [12] by

$$\Delta_1\varphi = g^{ij}\varphi_{,i}\varphi_{,j}. \tag{19}$$

The needed identity may now be obtained from Equation (18) as

$$4\pi\left(T^i_i - T^4_4\right) = \frac{\Delta_1\varphi}{V^2}, \tag{20}$$

which, for spherical coordinates, may be written as

$$4\pi\left(T^i_i - T^4_4\right) = \frac{g^{11}(\varphi_{,r})^2}{V^2}. \tag{21}$$

The total effective mass inside the 3-volume $dv'_3$ is then

*Charge, geometry, and effective mass*

$$M_{Eff}^{Out} = -\int_R (T_i^i - T_4^4) \sqrt{V}\, dv_3 = -\frac{1}{4}\int_R \frac{g^{11}(\phi,_r)^2}{\sqrt{V}}\, dv_3. \qquad (22)$$

Substitution of $V'$ from Eq. (11), along with $g^{11} = \left(1 + \frac{Q^2}{r^2}\right)$, $\phi = Q/r$, and $dv_3 = \frac{r^2}{\sqrt{V}} \sin\theta\, d\theta\, d\varphi$, yields

$$M_{Eff}^{Out} = -\int_R \frac{Q^2}{r^2}\, dr = \frac{Q^2}{R}. \qquad (23)$$

As expected, this is the same result as that given in Eq. 13.

**Summary**

The above results may then be summarized as in Eq. (14)

$$M_{Eff}^{In} + M_{Eff}^{Out} = m,$$

independent of the radius $R$. What this says is that the amount of "negative mass" due to the term $-Q^2/R$ in Eq. (9) is exactly compensated by the amount of "positive mass" contained in the region $r > R$. For $R$ infinite, $M_{Eff}^{In}$ is the Schwarzschild mass; and if $R < \infty$, $M_{Eff}^{In}$ is less than the Schwarzschild mass.

In their 1979 paper, Cohen and Gautreau [5] noted that: "As $R$ decreases, $M_T$ [here equal to $M_{Eff}^{In}$] also decreases because the electric field energy inside a sphere of radius $R$ decreases." And, one might add, as $R$ decreases, the field energy exterior to $R$ increases. This is equivalent to

*Charge, geometry, and effective mass*

$$M_{Eff}^{In} = \left(m - \frac{Q^2}{R}\right) \qquad \frac{dM_{Eff}^{In}}{dR} = \frac{Q^2}{R}$$

$$M_{Eff}^{Out} = \frac{Q^2}{R} \qquad \frac{dM_{Eff}^{Out}}{dR} = -\frac{Q^2}{R}, \qquad (24)$$

so that

$$\frac{dM_{Eff}^{In}}{dR} + \frac{dM_{Eff}^{Out}}{dR} = 0. \qquad (25)$$

While charge of either sign causes a negative curvature of space-time, the Einstein-Maxwell system of equations does not allow different geometric representations for positive and negative charges. This is a direct result of the fact that the sources of the Einstein-Maxwell system are embodied in the energy-momentum tensor, which depends only on the (non-gravitational) energy density—which is why charge enters as $Q^2$ above. Thus, a full geometrization of charge does not appear to be possible within the framework of the Einstein-Maxwell equations.

As mentioned earlier, no "realistic" sources for the Reissner-Nordström metric are considered in this paper. Realistic sources raise many interesting questions, among them are: Can a lone, charged black hole actually exist? If so, how can global charge neutrality be maintained?

**Acknowledgement**

I would like to thank Adam J. Marsh and Charles Nissim-Sabat for helpful conversations.

**REFERENCES**

[1] H. Reissner, "Über die Eigengravitation des elektrischen Feldes nach der Einstein'schen Theorie", *Ann. Physik*, **50**, 106-120 (1916); G. Nordström, "On the Energy of the Gravitational Field in Einstein's Theory", *Verhandl. Koninkl. Ned. Akad. Wetenschap., Afdel. Natuurk., Amsterdam* **26**, 1201-1208 (1918).

[2] S.W. Hawking and G.F.R. Ellis, "The Large Scale Structure of Space-Time" (Cambridge University Press, Cambridge 1973), pp. 156-161.

[3] The divergence of the Kretschmann scalar as $r \to 0$ indicates a real—as opposed to a coordinate dependent—singularity. It has been proposed that the Kretschmann scalar be called "the spacetime curvature" of a black hole; see: R.C. Henry, "Kretschmann Scalar for a Kerr-Neuman Black Hole", *Astrophys. J.* **535**, pp. 350-353 (2000).

[4] V. de la Cruz and W. Israel, "Gravitational bounce", *Nuovo Cimento* **51**, 744 (1967).

[5] J.M. Cohen and D.G. Gautreau, "Naked singularities, event horizon, and charged particles", *Phys. Rev. D* **19**, 2273-2279 (1979).

[6] W.A. Hiscock, "On the topology of charged spherical collapse", *J. Math. Phys.* **22**, 215 (1981).

[7] F. de Felice and C.J.S. Clarke, "Relativity on Curved Manifolds" (Cambridge University Press, Cambridge (1992), pp. 369-372.

[8] J.L. Synge, "Relativity: The General Theory" (North-Holland Publishing Company, Amsterdam 1966), Ch. VII, §5 and Ch. X, §4.

[9] R. Gautreau and R.B. Hoffman, "The Structure of the Sources of Weyl-Type Electrovac Fields in General Relativity", *Il Nuovo Cimento* **16**, 162-171 (1973).

[10] E.T. Whittaker, "On Gauss'theorem and the concept of mass in general relativity," *Proc. Roy. Soc*. London **A149**, 384 (1935).

[11] R. Arnowitt, S. Deser, and C. W. Misner, "The Dynamics of General Relativity", contained in: L. Witten (Editor), "Gravitation: an introduction to current research" (John Wiley & Sons, Inc., New York 1962), pp. 227-265.

[12] L.P. Eisenhart, "Riemannian Geometry", (Princeton University Press, Princeton 1997), p.41.